\documentclass[twocolumn,secnumarabic,amssymb, amsmath, nofootinbib,tightenlines, nobibnotes, aps, prl,epsfig]{revtex4}
\usepackage{graphicx}
\usepackage{dcolumn}
\usepackage{bm}
\begin{document}
\preprint{APS/123-QED}
\title{ Nonlinear terms from GLR-MQ evolution}

\author{G.R.Boroun}%
 \email{boroun@razi.ac.ir }

\affiliation{Department of  Physics, Razi University, Kermanshah
67149, Iran}

\date{\today}
\begin{abstract}
We show that the saturation exponent is more effective than the
hard pomeron exponent in the nonlinear
 terms for the GLR-MQ evolution equations. For the gluon distribution the nonlinear terms are found to play an increasingly important
 role than the singlet distribution. The nonlinear corrections for the singlet and gluon distributions are
 important in a limited range of $Q^2$ and are independent of the upward and downward
 evolution. This allows us to single out the region where the nonlinear corrections
  may provide evidence for recombination. We numerically study the
  nonlinear terms for nuclear singlet and gluon distributions and
  quantify the impact of gluon recombination in a wide range of
  $Q^2$. This analysis indicates that the nonlinear effects are
  most pronounced for heavy nuclei and are sizable at
  $2{\lesssim}Q^2{\lesssim}20~\mathrm{GeV}^2$ for light and heavy
  nuclei.\\
\end{abstract}
 \pacs{***}
\maketitle
On the basis of the GLR-MQ equations [1,2] it is known that the
dominant source for the nonlinear corrections at low $x$ is the
conversion of the two gluon ladders merge into a gluon or a
quark-antiquark pair\footnote{The second process is the
Glauber-like rescattering of the quark-antiquark fluctuations off
gluons.}. This fact makes it possible to relate the evolution of
the parton distributions directly to the gluon-gluon fusion terms
as \footnote{For future discussion please see Appendix.}
\begin{eqnarray}
\Delta{xg(x,Q^2)}=-\frac{81}{16}\frac{\alpha_{s}^{2}(Q^{2})}{\mathcal{R}^{2}Q^{2}}\int_{\chi}^{1}\frac{dz}{z}[\frac{x}{z}g(\frac{x}{z},Q^{2})]^{2}
\end{eqnarray}
 and
\begin{eqnarray}
\Delta{xq_{s}(x,Q^2)}=-\frac{27\alpha_{s}^{2}(Q^{2})}{160\mathcal{R}^{2}Q^{2}}[xg(x,Q^{2})]^{2},
\end{eqnarray}
where
\begin{eqnarray}
\Delta{xf_{i}(x,Q^2)}=\frac{\partial{xf_{i}(x,Q^{2})}}{\partial{\ln}Q^{2}}-\frac{\partial{xf_{i}(x,Q^{2})}}{\partial{\ln}Q^{2}}|_{\mathrm{DGLAP}}
\end{eqnarray}
Here $\mathcal{R}$ is the characteristic radius of the gluon
distribution in the hadronic target and $\chi=\frac{x}{x_{0}}$,
where $x_{0}$ is the boundary condition that the gluon
distribution joints smoothly onto the linear region. The ratio of
$\frac{\Delta{xg(x,Q^2)}}{\Delta{xq_{s}(x,Q^2)}}$ is independent
of $\mathcal{R}$ and the running coupling as we can write
\begin{eqnarray}
r_{gs}=\frac{\Delta{xg(x,Q^2)}}{\Delta{xq_{s}(x,Q^2)}}=\frac{30}{[xg(x,Q^{2})]^{2}}\int_{\chi}^{1}\frac{dz}{z}[\frac{x}{z}g(\frac{x}{z},Q^{2})]^{2}.
\end{eqnarray}
The gluon distribution is restricted to the Regge-like behavior.
The small $x$ behavior of the gluon distribution is given by
\begin{eqnarray}
xg(x,Q^2)=f_{g}(Q^2)x^{-\lambda_{g}},
\end{eqnarray}
where it is hard-pomeron dominated if
$\lambda_{g}=\epsilon_{0}{\simeq}0.5$ [3,4]. This more singular
behavior obtained by solving the Lipatov equation, together with
shadowing effects which can restrict the small-$x$ growth of the
more singular distributions [5]. This value in the saturation
scale $Q_{s}$ of the dipole cross section of the GBW model [6] is
approximately ${\simeq}0.3$. As a consequence, the power
$\lambda_{g}$ which governs change of the gluon distribution with
$x$ in the Regge and saturation models, is independent of $Q^2$.
These models predict a single singularity in (5) at small-$x$.
Substituting the gluon distribution in (1) and (4) and then
integrating, one gets
\begin{eqnarray}
\Delta{xg(x,Q^2)}=-\frac{81}{16}\frac{\alpha_{s}^{2}(Q^{2})}{\mathcal{R}^{2}Q^{2}}[xg(x,Q^{2})]^{2}\int_{\chi}^{1}{dz}z^{2\lambda_{g}-1}
\end{eqnarray}
and
\begin{eqnarray}
r_{gs}={30}\int_{\chi}^{1}{dz}z^{2\lambda_{g}-1}.
\end{eqnarray}
The ratio $r_{gs}$ is dependent on the  pomeron
intercept\footnote{$\lambda_{g}$ is the pomeron intercept minus
one.}. The formal expression for this ratio is fully known to the
exponents. The relative importance of the saturation exponent in
(7) to the hard pomeron is shown in Fig.1 in a wide range of $x$.
This ratio is independent of $Q^2$ and shows that the nonlinear
terms in the evolution equations are more effective with the
saturation exponents. For instance, the value of $r_{gs}$ with the
saturation exponent at $x=10^{-5}$ is increased by 55$\%$ compared
to the hard pomeron exponent. This ratio saturates at
$x{\lesssim}10^{-3}$ for $\lambda_{g}=0.452$ and
at $x{\lesssim}10^{-4}$ for $\lambda_{g}=0.288$.\\
\begin{figure}
\centerline{
\includegraphics[width=0.5\textwidth]{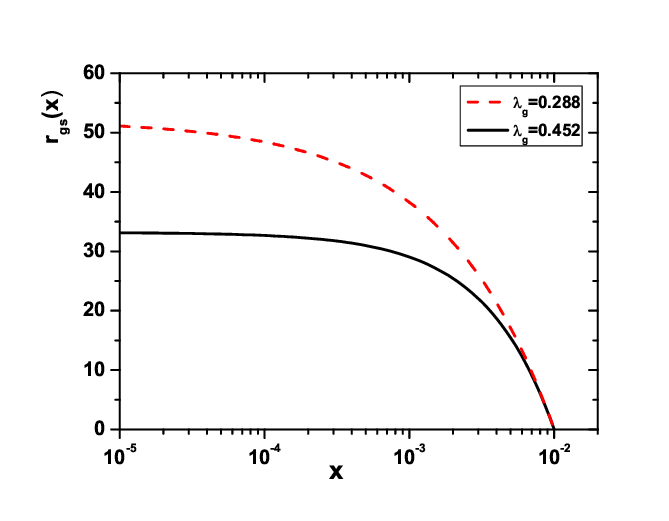}}
\caption{ Effect of exponents. The dashed line is the ratio of
$r_{gs}$ with the saturation exponent and the solid line is with
the hard pomeron exponent.}\label{Fig1}
\end{figure}
In Ref.[7], the author investigated how the gluon distribution is
obtained from the dipole fits where the exponent lambda is defined
with the fixed value $\lambda=0.28$. The author has defined the
leading twist relationship between the unintegrated and integrated
gluon distribution,
$g(x,Q^2)=\int_{0}^{Q^2}\frac{dk^2}{k^2}f(x,k^2)$ in Ref.[7]. This
distribution is dependent on the saturation exponent. The
nonlinear terms in the GLR-MQ evolution equations with the gluon
distribution [7] help to estimate the effect of the gluon
recombination independent of the running coupling, while most of
existing predictions are done in the leading and next-to-leading
logarithmic approximations. Indeed, the gluon distribution of Ref.
[7] is used as initial condition to compute the numerical solution
of GLR-MQ equations. The gluon and singlet nonlinear terms (i.e.,
Eqs.(6) and (2) respectively) are shown in Fig.2 for
$Q_{0}^{2}=1.69~\mathrm{GeV}^2$ at the exponent
$\lambda_{g}=0.288$. Figure 2 presents the results of the
nonlinear terms in the GLR-MQ evolution equations for the singlet
and gluon distributions at $R=5~\mathrm{GeV}^{-1}$ and
$2~\mathrm{GeV}^{-1}$ at $Q_{0}^2=1.69~\mathrm{GeV}^2$ as a
function of the momentum fraction $x$.
\begin{figure}
\centerline{
\includegraphics[width=0.55\textwidth]{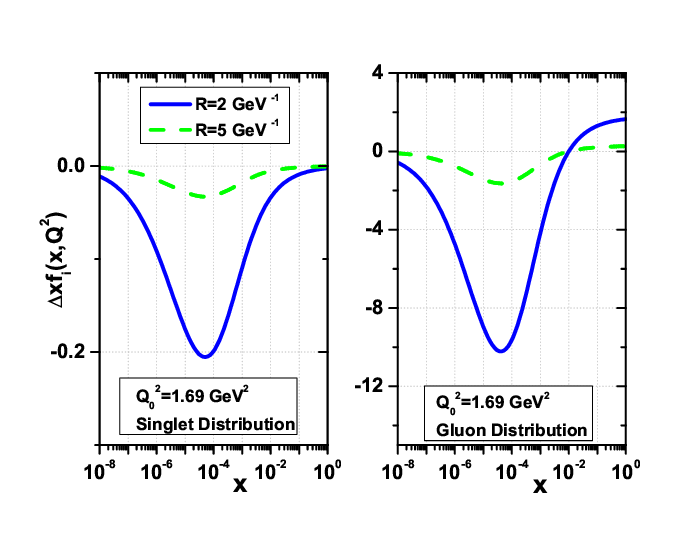}}
\caption{ Results of the nonlinear terms for the singlet (left)
and gluon (right) distributions at $R=5~\mathrm{GeV}^{-1}$ (dashed
curves) and at $R=2~\mathrm{GeV}^{-1}$ (solid curves) at
$Q_{0}^2=1.69~\mathrm{GeV}^2$.}\label{Fig2}
\end{figure}
A gaussian behavior for $\Delta{xf_{i}}$, due to the gluon
distribution, is predicted. We also found a symmetry between the
regions of large and small $x$ for the singlet and gluon
distributions, which is illustrated in Figs.2 and 3. In these
figures (i.e., Figs.2 and 3), we note that the size parameter $R$
is important on magnitude of the recombination terms [8-10]. It is
clear that the gluon recombination terms are enlarged by a factor
of almost 50 compared to the singlet distribution. The signal of
the gluon recombination terms in the singlet and gluon
distributions is a decrease in the scaling violation as compared
to  standard evolution and it is independent of the upward and
downward evolutions [11].
\begin{figure}
\centerline{
\includegraphics[width=0.55\textwidth]{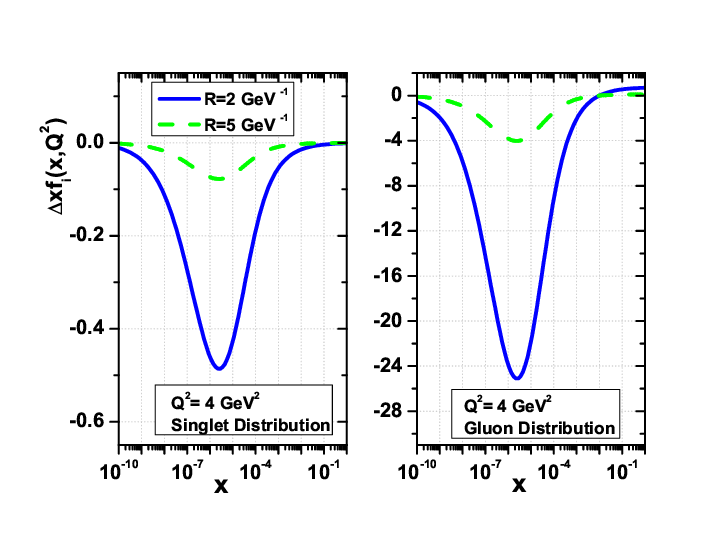}}
\caption{ The same as Fig.2 for
$Q^2=4~\mathrm{GeV}^2$.}\label{Fig3}
\end{figure}
Minimum  $\Delta{xf_{i}}$ values in the singlet and gluon
distributions at $Q_{0}^{2}=1.69~\mathrm{GeV}^2$ and
$Q^2=4~\mathrm{GeV}^2$ are obtained in the range $10^{-5}
{\lesssim}x{\lesssim}10^{-4}$ and $10^{-6}
{\lesssim}x{\lesssim}10^{-5}$  in Figs.2 and 3, respectively. The
effective values of $\Delta{xf_{i}}$ are visible from these
figures and show the nonlinear corrections in a wide range of
$x$.\\
In order to clarify the results from Figs.2 and 3, we computed
contributions of the nonlinear terms for $\Delta{xf_{i}}$ in a
wide range of $Q^2$ in Figs.4 and 5. It should be noted that the
optimal values for these corrections are in the ranges
$10^{-6}{\lesssim}x{\lesssim}10^{-3}$ and
$2{\lesssim}Q^2{\lesssim}10~\mathrm{GeV}^2$. The results of this
analysis show that the nonlinear corrections increase with
increases of $Q^2$ and decreases of $x$ in the defined range.
Indeed, the gluon recombination plays an essential role in gluon
dynamics at low $x$ and $Q^2$ values, since it gives such a
radically different expectation on the behavior of the singlet and
gluon distribution functions.\\
\begin{figure}
\centerline{
\includegraphics[width=0.55\textwidth]{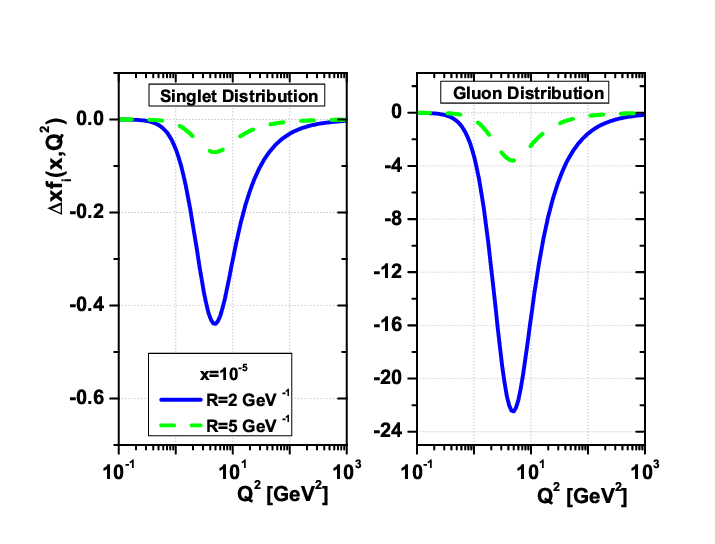}}
\caption{ Results of the nonlinear terms for the singlet (left)
and gluon (right) distributions at $R=5~\mathrm{GeV}^{-1}$ (dashed
curves) and at $R=2~\mathrm{GeV}^{-1}$ (solid curves) at
$x=10^{-5}$ in a wide range of $Q^2$.}\label{Fig4}
\end{figure}
\begin{figure}
\centerline{
\includegraphics[width=0.55\textwidth]{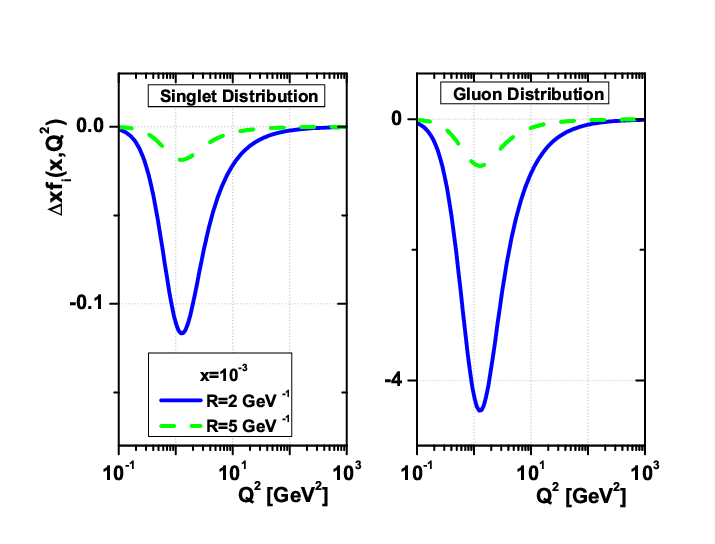}}
\caption{ The same as Fig.4 for $x=10^{-3}$.}\label{Fig5}
\end{figure}
The results that are obtained in the above (i.e., Figs.2-5) can be
confirmed by considering the general behavior for the gluon
distribution (i.e., without the Regge -like behavior). In the
following, we consider this behavior for the nuclear gluon
density. Let us recall that the nonlinear terms for nuclear parton
distribution functions are related to the nuclear gluon
distribution by the nonlinear evolution equations. Indeed, the
GLR-MQ equations remain a good approximation at intermediate
energy where the $2{\rightarrow}1$ gluon fusion dominates, and
should be sufficient for simulating the events in $eA$ collisions
at the large hadron electron collider (LHeC) [12] and the
Electron-Ion Collider (EIC) [13] energies. The nonlinear
corrections in $eA$ processes satisfy the GLR-MQ evolution
equations, as the nonlinear terms for evolution of the nuclear
PDFs (gluon and singlet) reads [14]
\begin{eqnarray}
\Delta{xg_{A}(x,Q^2)}=-\frac{81}{16}\frac{A\alpha_{s}^{2}(Q^{2})}{\mathcal{R}^{2}_{A}Q^{2}}\int_{\chi}^{1}\frac{dz}{z}[\frac{x}{z}g_{A}(\frac{x}{z},Q^{2})]^{2}
\end{eqnarray}
 and
\begin{eqnarray}
\Delta{xq_{s,A}(x,Q^2)}=-\frac{27A\alpha_{s}^{2}(Q^{2})}{160\mathcal{R}^{2}_{A}Q^{2}}[xg_{A}(x,Q^{2})]^{2}.
\end{eqnarray}
For a nuclear target with the mass number A, the nuclear radius
$\mathcal{R}_{A}=2~\mathrm{GeV}^{-1}{\times}A^{1/3}$ is based on
how the gluons have a hotspot-like structure.\\
The two-gluon distribution [10], in the leading twist
approximation, is taken to be the product of two conventional
one-gluon distribution as $
xg^{(2)}{\sim}\frac{1}{\pi\mathcal{R}_{p}^{2}} [xg(x,Q^2)]^2 $.
For a nuclear target with the mass number $A$, the nuclear parton
distribution functions scale approximately as $A$ [16]. The
two-gluon distribution for nuclei in Eqs.(8), (9) is defined to be
$xg^{(2)}_{A}(x,Q^2)=\frac{A}{\pi
\mathcal{R}_{A}^{2}}[xg_{A}(x,Q^2)]^2$, where $g_{A}(x,Q^2)$ is
obtained from Ref.[7] with replacement
$Q_{s}^{2}{\rightarrow}Q_{s}^{2A}$
($Q_{s}^{2}=Q_{0}^{2}(x/x_{0})^{-\lambda}$ where $Q_{0}$ and
$x_{0}$ are free parameters), where $Q_{s}^{2A}=A^{1/3}Q_{s}^{2}$
[15]. In order to estimate the strength of nonlinear terms in eA
processes, we can compute the nonlinear corrections to the gluon
and singlet evolutions using Eqs.(8) and (9) respectively. The net
effect of the nonlinear terms of the singlet and gluon
distributions for C-12 and Au-197 are shown in Figs.6 and 7 for
$x=10^{-5}$ and $10^{-3}$ in a wide range of $Q^2$, respectively.
These results are comparable with the difference between the
linear and nonlinear data in Ref.[16]
which is based on nCTEQ15 nPDFs for Au-197.\\
\begin{figure}
\centerline{
\includegraphics[width=0.55\textwidth]{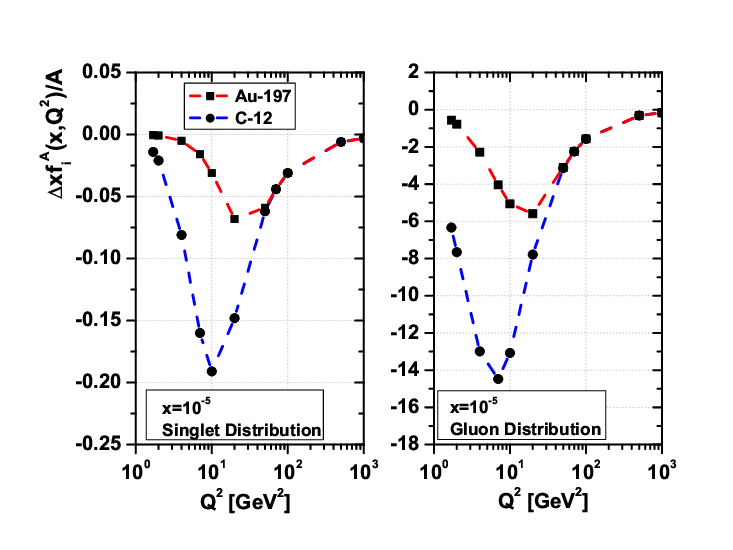}}
\caption{ Results of the nonlinear terms for the nuclear singlet
(left) and gluon (right) distributions for C-12 (dashed-circles )
and Au-197 (dashed-squares) at $x=10^{-5}$ in a wide range of
$Q^2$.}\label{Fig6}
\end{figure}
Minimum  $\frac{1}{A}{\Delta}xf^{A}_{i}$ values in the singlet and
gluon distributions for C-12 and Au-197 at $x=10^{-5}$ and
$x=10^{-3}$ are obtained in the range
$10{\lesssim}Q^2{\lesssim}20~\mathrm{GeV}^2$ and
$2{\lesssim}Q^2{\lesssim}10~\mathrm{GeV}^2$  in Figs.6 and 7,
respectively. The effective values of
$\frac{1}{A}{\Delta}xf^{A}_{i}$ can be seen  from these figures
and show the nonlinear corrections in a wide range of nuclei. The
minimum value increases with the increase of mass number $A$ and
decreases with its decrease. This value shifts to larger values of
$Q^2$ as the mass number $A$ increases. It is clear that the
nonlinear terms are very small as $Q^2$ increases and this
behavior is independence of the mass number $A$. The deviation of
these terms from zero shows the importance of nonlinear
effects on light and heavy nuclei.\\
\begin{figure}
\centerline{
\includegraphics[width=0.55\textwidth]{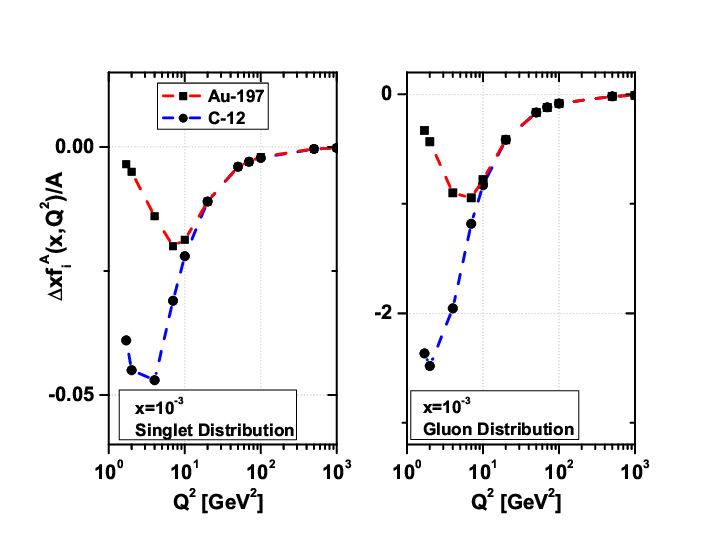}}
\caption{ The same as Fig.6 for $x=10^{-3}$.}\label{Fig7}
\end{figure}
In summary, we have used the gluon distribution that is defined in
Ref. [7] for the nonlinear terms in the GLR-MQ evolution
equations. The difference between the GLR-MQ and DGLAP equations,
confirmed the importance of the nonlinear corrections for small
$x$ and $Q^2$, whose magnitude increases with a decrease of $x$
and an increase of the atomic number $A$ in a limited range of
$Q^2$. The nonlinear effect is much stronger for the singlet and
gluon distributions in a wide range of $x$ and nuclei at
$2{\lesssim}Q^2{\lesssim}20~\mathrm{GeV}^2$. We observe that the
ratio ${\Delta}xg_{A}/{\Delta}xq_{s,A}$ is suppressed by
$\approx50\%$ for a free proton and increases for nuclei with
increase of the mass number A. Having checked that these ratios
give a good description of the gluon-gluon recombinations rather
than the gluon-quark splitting in a future electron-ion collider.
Indeed, the nonlinear terms  decrease the results of the evolution
of parton distributions and they are independent
 of the upward and downward evolution.\\

\subsection{Appendix}

The nonlinear evolution equations relevant at high gluon densities
have been studied at small $x$ in Ref.[17], where we expect
annihilation or recombination of gluons to occur. A measurement of
$g(x,Q^2)$ in this region probes a gluon of transverse size $\sim
1/Q$, therefore the transverse area of the thin disc that they
occupy is $\sim xg(x,Q^2)/Q^2$. The nonlinear effects, at
sufficiently small $x$ where $W{\lesssim}\alpha_{s}$, can be
calculated in perturbative QCD. Here
$W{\sim}\frac{\alpha_{s}(Q^2)}{\pi{\mathcal{R}^2}Q^2}xg(x,Q^2)$
and $\pi{\mathcal{R}^2}$ is the transverse area where
$\mathcal{R}$ is the proton radius. The QCD evolution equation
modified for the gluon distribution is defined by the following
form
\begin{eqnarray}
\frac{\partial{xg(x,Q^{2})}}{\partial{\ln}Q^{2}}&=&P_{gg}{\otimes}g+P_{gq}{\otimes}q
-\frac{81}{16}\frac{\alpha_{s}^{2}(Q^{2})}{\mathcal{R}^{2}Q^{2}}\theta(x_{0}-x)\nonumber\\
&&{\times}\int_{x}^{x_{0}}\frac{dx'}{x'}[{x'}g({x'},Q^{2})]^{2},
\end{eqnarray}
where the $\theta$ function reflects the ordering in longitudinal
momenta as for $x{\geq}x_{0}$ the shadowing correction is
negligible ($x_{0}{=}10^{-2}$).  The shadowing term has a minus
sign  because the scattering amplitude corresponding to the gluon
ladder is predominantly imaginary. There are also shadowing
corrections to the evolution equation for the sea-quark
distributions as
\begin{eqnarray}
\frac{\partial{xq_{s}(x,Q^{2})}}{\partial{\ln}Q^{2}}&=&P_{qg}{\otimes}g+P_{qq}{\otimes}q_{s}
-\frac{27\alpha_{s}^{2}(Q^{2})}{160\mathcal{R}^{2}Q^{2}}[xg(x,Q^{2})]^{2}\nonumber\\
&&+\mathrm{G_{HT}},
\end{eqnarray}
where the higher dimensional gluon term $\mathrm{G_{HT}}$ is here
assumed
to be zero.\\


\subsection{ACKNOWLEDGMENTS}

We are grateful to the Razi University for financial support of
this project.\\

\section{References}
1. L. V. Gribov, E. M. Levin and M. G. Ryskin, Phys. Rept.
{\bf100}, 1 (1983).\\
2. A. H. Mueller and J. w. Qiu, Nucl. Phys. B {\bf268}, 427
(1986).\\
3. A.Donnachie and P.V.Landshoff, Phys.Lett.B {\bf533}, 277
(2002).\\
4. P.V.Landshoff, arXiv [hep-ph]/0203084.\\
5. L.H.Orr and W.J.Stirling, Phys.Rev.Lett. {66}, 1673 (1991).\\
6. K.Golec-Biernat and  M.W$\ddot{\mathrm{u}}$sthoff, Phys. Rev. D
{\bf59},  014017 (1998).\\
7. R.S.Thorne, Phys.Rev.D {\bf71}, 054024 (2005).\\
8. K.Prytz, Eur.Phys.J.C {\bf22}, 317 (2001).\\
9. K.J.Eskola et al., Nucl.Phys.B {\bf660}, 211 (2003).\\
10. W.Zhu, Phys.Lett.B {\bf389}, 374 (1996).\\
11. Y.Shi, Shu-Yi Wei and J.Zhou, Phys.Rev.D {\bf107}, 016017
(2023).\\
12. LHeC Collaboration, FCC-he Study Group, P. Agostini, et al.,
J. Phys. G, Nucl. Part. Phys. {\bf48},  110501 (2021).\\
13. R.Abdul Khalek et al., Nucl. Phys.A {\bf 1026}, 122447
(2022).\\
14. K.J.Eskola et al., Phys.Lett.B {\bf532}, 222 (2002).\\
15. F.Carvalho, F.O.Duraes, F.S.Navarra and S.Szpigel, Phys.Rev.C
{\bf79}, 035211 (2009).\\
16. J.Rausch, V.Guzey and M.Klasen, Phys.Rev.D {\bf107}, 054003
(2023).\\
17. J.Kwiecinski et al., Phys.Rev.D {\bf42}, 3645 (1990).\\




\end{document}